\documentclass[conference,compsoc]{IEEEtran}

\usepackage[T1]{fontenc}
\usepackage{booktabs}
\usepackage{hyperref}
\hypersetup{hidelinks}
\usepackage{enumitem}
\usepackage{graphicx}
\usepackage{float}
\usepackage{subcaption}
\usepackage{array}
\usepackage{amsmath}
\usepackage{amssymb}
\usepackage{microtype}
\usepackage{algorithm}
\usepackage{algpseudocode}
\usepackage{multirow}
\usepackage[table]{xcolor}
\usepackage[most]{tcolorbox}
\definecolor{rescored}{RGB}{214,234,248}
\definecolor{findingbg}{RGB}{240,244,248}
\definecolor{findingrule}{RGB}{150,165,180}
\definecolor{sevcrit}{RGB}{238,238,234}
\definecolor{promptbg}{RGB}{247,249,251}
\definecolor{promptrule}{RGB}{150,165,180}
\definecolor{prompthead}{RGB}{42,84,124}
\definecolor{promptelide}{RGB}{120,128,136}
\newtcolorbox{promptbox}[1][]{colback=promptbg, colframe=promptrule,
  boxrule=0.5pt, arc=1.5pt, boxsep=1pt, left=5pt, right=5pt, top=3pt, bottom=3pt,
  fonttitle=\bfseries\scriptsize, coltitle=white, colbacktitle=promptrule,
  title={#1}, breakable}
\newcommand{\phead}[1]{\textcolor{prompthead}{\bfseries\texttt{\#\# #1}}}
\newcommand{\pelide}[1]{\textcolor{promptelide}{\normalfont\itshape #1}}

\newtcolorbox{finding}{colback=findingbg, colframe=findingrule,
  boxrule=0.4pt, arc=1.5pt, boxsep=1pt, left=4pt, right=4pt,
  top=2pt, bottom=2pt, before skip=4pt, after skip=4pt}

\begin{document}

\title{Ranked by the Matcher: A Reproducibility Audit of Knowledge Graph Extraction from Threat Reports}

\author{
\IEEEauthorblockN{Safayat Bin Hakim}
\IEEEauthorblockA{\textit{University of Maryland, Baltimore County}\\
Baltimore, Maryland, USA\\
shakim3@umbc.edu}
\and
\IEEEauthorblockN{Houbing Herbert Song}
\IEEEauthorblockA{\textit{University of Maryland, Baltimore County}\\
Baltimore, Maryland, USA\\
songh@umbc.edu}}

\maketitle
\IEEEpeerreviewmaketitle

\begin{abstract}
Security teams and researchers choose knowledge-graph extraction tooling for threat reports on the strength of published triple-F1 scores, yet those scores depend on how predicted triples are matched to gold annotations. We could reimplement the stated matching rule for only five of twelve inspected systems. Re-scoring ten system outputs on shared documents under eight protocols reverses eleven of forty-five pairwise orderings; one fixed prediction set spans 0.16--0.70 F1. On an external, human-adjudicated set, no mechanical matcher---lexical, embedding, or entailment---agrees with the reviewers more than seven times in ten; an LLM judge agrees far more often. To separate component effects from matcher rewards, we build CTIForge, whose deterministic validation layer can vary while extraction is held byte-identical. Across seven tested deployment configurations, no hosted backbone loses precision under validation and every offline one does. Backbone, decoding, and prompting covary across those configurations; on the one backbone we could serve both ways, serving alone reproduces the split. It coincides with a roughly 2.8-fold increase in actions explicitly disputing entity type, consistent with hand-written rules encoding the conventions of the extractor against which they were developed. We release the pipeline, protocol suite, and per-triple audit records.
\end{abstract}


\section{Introduction}
\vspace{-2mm}

Cyber threat intelligence (CTI) is distributed as narrative prose in vendor reports, and converting it into knowledge graphs of threat actors, malware, vulnerabilities, techniques and infrastructure is now routine in security operations~\cite{sun2023cti,wagner2019cti}. A steady stream of extraction systems has followed, most recently built on large language models. Practitioners and researchers pick among them the way one picks among any tools: by reading the numbers each paper reports.

The problem is not simply that extraction is imperfect; it is that the scores used to compare imperfect extractions may be incomparable.

Those numbers are almost always triple F1 against gold annotations---the harmonic mean of how many predicted (subject, relation, object) statements match an annotated one and how many annotated ones are recovered. What that number means depends entirely on when a predicted triple is deemed to match an annotated one. Suppose a system predicts (APT29, \emph{uses}, Cobalt Strike) where the annotation reads (APT29, \emph{deployed}, Cobalt Strike Beacon). Whether that counts as correct is not a property of the prediction but a decision: whether entity names must be identical or merely overlap, whether relation labels must agree exactly or only be compatible, whether a reversed subject and object counts, and whether matching happens within a document or across a pooled corpus.

We audited whether the matching rules of twelve recent CTI extraction systems and benchmarks can be reconstructed from their publications (Table~\ref{tab:protocol-survey}). No two inspected papers specify the same rule. Five rules were reimplementable, three named a family without fixing it, and four stated no criterion. Fieblinger et al.~\cite{fieblinger2024actionable}, for example, reject exact-match evaluation for generative output and adopt $n$-gram overlap.

The most consequential case is CTI-Nexus~\cite{cheng2025ctinexus}, a frequent comparison target. It counts a prediction as correct when its elements ``semantically match'' a gold triplet, but does not specify the encoder, threshold, or decision rule. A later paper repeating that comparison without defining its own matcher inherits a criterion it cannot reproduce.

\textbf{This paper asks how much the matcher decides.} Re-scoring ten published systems on a shared document set under eight protocols reverses \textbf{eleven of forty-five} pairwise orderings. Holding one fixed prediction set constant and varying nothing but the matcher moves its triplet F1 from \textbf{0.16 to 0.70}. On the external 378-item GRID calibration set~\cite{huang2026grid}, no mechanical protocol agrees with multi-reviewer human judgment on more than 71\% of items. Pooling a corpus before matching---which several public harnesses do, including, until we found it, our own---credits a prediction from one report against a different report's gold and inflates true positives by 6.4\%.

The security consequence is direct. A defender selecting an extraction pipeline, or a researcher choosing a baseline to improve on, is acting on evidence that may not survive a change of scoring rule. The observed protocol spread can make identical predictions on identical documents look like outputs from different systems.

\begin{figure*}[!t]
\centering
\includegraphics[width=0.85\textwidth]{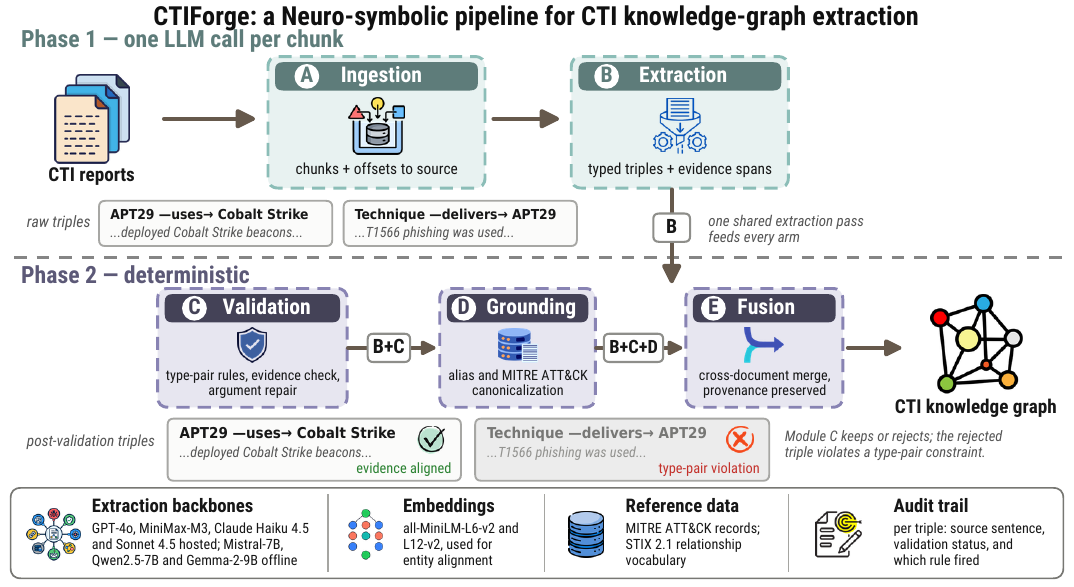}
\caption{CTIForge pipeline architecture. Extraction is a single LLM call per chunk; every stage after it is deterministic, which is what allows all ablation arms to be derived from one extraction pass.}
\label{fig:system-overview}
\vspace{-2mm}
\end{figure*}

\textbf{A question published artifacts cannot answer.} Auditing released outputs bounds what comparisons mean, but it cannot say what any individual design choice contributes, because the choice cannot be switched off in someone else's artifact. We therefore build \textbf{CTIForge}, a five-stage neuro-symbolic pipeline whose deterministic validation layer---type-pair constraints, indicator normalization, evidence-alignment filtering, with every action logged per triple---can be enabled or disabled while extraction is held byte-identical (Fig.~\ref{fig:system-overview}). CTIForge is apparatus for this paper rather than its proposal, and it is released as such.

Used that way, CTIForge yields an unexpected result. Across seven tested configurations, validation's precision effect \textbf{changes sign}: no hosted backbone loses precision and every offline one does, with no overlap. This grouping is observational: serving mode is not varied independently of backbone, decoding, or prompting.

The split is not ordered by unvalidated F1 or listed parameter count. Instead, it tracks how much of the validator's work disputes entity \emph{types}: loose typing can trigger constraints on triples that are substantively correct. Because the rules were developed against frontier-model output, they encode its conventions. Deterministic does not imply model-agnostic, in this pipeline or in neuro-symbolic pipelines generally. Canonicalization, meanwhile, rewrites a small share of triples that the metric does not register.

This paper makes four contributions. First, we provide a \textbf{reproducibility audit} that re-scores ten system outputs on shared documents under a common protocol suite. Second, we \textbf{calibrate mechanical protocols} against an external multi-reviewer set rather than assuming that semantic similarity tracks judgment. Third, we give a \textbf{shared-extraction ablation} of deterministic validation and canonicalization across seven deployment configurations, isolating each downstream module's effect conditional on fixed predictions. Fourth, we \textbf{measure two correctable evaluation defects}: corpus-pooled matching and comparative use of metrics defined by a participant's own schema. We release the code, configurations, logs, and per-triple audit records (Appendix~\ref{sec:appendix-data}).

The implementation, the evaluation harnesses, and the raw error-taxonomy logs underlying Section~\ref{sec:error-dist} are publicly available. The per-triple validation records reported here can be inspected directly.

\vspace{-2mm}
\section{Related Work}
\vspace{-2mm}

\vspace{-1mm}

\subsection{Evaluating CTI Extraction}

\vspace{-2mm}

Benchmarks for CTI language tasks have grown quickly---CTIBench~\cite{alam2024ctibench} and SEvenLLM~\cite{ji2024sevenllm} both evaluate LLMs across CTI abilities---but graph-construction results still rely almost exclusively on triple F1 against gold annotations. Because the matching rule is often informal or absent, published figures are not directly comparable. Table~\ref{tab:protocol-survey} groups twelve systems and benchmarks by how completely we could reconstruct that rule: five were reimplementable from the paper, three left an encoder or threshold unfixed, and four stated no criterion. Seven publications therefore leave unresolved choices about name normalization, relation compatibility, or document-level scoping.

GRID~\cite{huang2026grid} is the closest prior work to treat the instrument itself as an evaluation object, calibrating an LLM judge against multi-reviewer human adjudication. We use that calibration set to evaluate ten mechanical protocols and extend the question to ordering sensitivity across eight protocols and ten system outputs. We classified each publication by reading its evaluation section and searching for a definition of when a prediction counts as correct; where none was found, we record that rather than infer one. This was one coder and one pass, with no inter-rater statistic~\cite{olszewski2023get}; criteria stated only in released code may have been missed. The counts characterize these inspected publications, not all CTI extraction systems.

\textbf{Re-evaluation as a contribution.} Security research has an established line of work that treats evaluation practice as the object of study rather than the means. Olszewski et al.~\cite{olszewski2023get} measure the problem at field scale, finding across nearly 750 machine-learning security papers that 60\% publish no code to run their experiments; Arp et al.~\cite{arp2022dosdonts} catalog recurring pitfalls that inflate reported performance; and TESSERACT~\cite{pendlebury2019tesseract} shows that removing temporal bias changes malware-classification conclusions. Narrower re-evaluations find that graph-based intrusion-detection results largely fail to reproduce~\cite{wang2025gidsreeval}, dataset duplication and untuned baselines explain much of the reported advantage in static Android malware detection~\cite{alam2025androidreeval}, and most of 38 surveyed DGA papers rest on a fragile practicality basis~\cite{cebere2024dga}. In CTI, Schlette et al.~\cite{schlette2021measuring} define and visualize quality dimensions, Yang et al.~\cite{yang2025automated} assess feed trustworthiness, and Geras and Schreck~\cite{geras2024bigbeast} examine operational quality assurance. That work asks whether the intelligence is good; we ask whether the numbers used to compare the systems that build it are.

The same instrument problem appears outside security. Wadhwa et al.~\cite{wadhwa2023revisiting} abandon exact matching for human evaluation because generative models express correct relations in forms the metric rejects. Swarup et al.~\cite{swarup2025llm4re} similarly find that extraction quality is not robust to the data encountered, while GenRES~\cite{jiang2024genres} proposes a multidimensional alternative to precision and recall against annotated references. Tan et al.~\cite{tan2022revisiting} locate a complementary failure in the gold data: re-annotating 4{,}053 DocRED documents recovers relations the original scheme missed. SQC-Score~\cite{fan2024evaluating} addresses both problems by refining the match and enriching the gold set with natural-language inference. We test entailment in the opposite direction---as a matcher against existing gold rather than a means of extending it---and find that it adds no value there (Section~\ref{sec:protocol}); this does not bear on the enrichment task. Flood et al.~\cite{flood2024bad} likewise show in security that auditing an evaluation artifact can itself be a contribution.

We contribute the CTI knowledge-graph case, where the bias is neither temporal nor sampling-related nor a property of the data, but resides in the matching rule. Concurrent work on ATT\&CK technique extraction reports a similar sensitivity to evaluation setup~\cite{ryan2026attackclass,haque2026beyondsingle}. The defect is an artifact of reporting, not of experimental design, and therefore correctable at no cost.

\subsection{CTI Extraction Systems}

\vspace{-2mm}

Extraction from threat reports has moved through three generations, surveyed in depth by Rahman et al.~\cite{rahman2023attackers} and, for relation extraction, systematized by Arikkat et al.~\cite{arikkat2024relation}. Those works catalog the methods; we audit the measurement used to rank them. Annotated corpora such as MalwareTextDB~\cite{lim2017malwaretextdb} established the gold-set convention used by later systems. Rule-based systems such as TTPDrill~\cite{husari2017ttpd} offered interpretability without adaptability. Supervised neural pipelines---TIMiner~\cite{zhao2020timiner}, Open-CyKG~\cite{sarhan2021open}, CyberRel~\cite{guo2021cyberrel}, Vulcan~\cite{jo2022vulcan}, CSKG4APT~\cite{ren2022cskg4apt}, KnowCTI~\cite{wang2024knowcti}, CyberEntRel~\cite{ahmed2024cyberentrel}, and AttacKG+~\cite{zhang2025attackkg}---improved coverage at the cost of annotation dependence and brittleness to new phrasing.

In-context learning removed the annotation requirement: CTI-Nexus~\cite{cheng2025ctinexus} uses demonstration retrieval and multi-phase prompting, LLM-TIKG~\cite{hu2024llmtikg} distills a smaller model from LLM-labeled data, and CTIKG~\cite{huang2024ctikg} organizes multiple agents over long reports. A parallel line adds structure around the model through role specialization~\cite{zhong2026ctiexpert}, ontology conformance~\cite{kim2026anchor,bouchiha2026tactickg,mouiche2026ctikg}, or explicit reasoning traces~\cite{yang2026ctithinker}. Outside security, EDC~\cite{zhang2024extract} makes canonicalization a named stage of LLM-based graph construction, as we do in Module~D.

What these systems share is the reporting convention this paper audits. Each is evaluated by matching predicted triples against gold annotations, each reports triple F1 as the headline, and---per Table~\ref{tab:protocol-survey}---most do not fix the rule that decides a match. A unified re-evaluation of TTP extraction reports the same difficulty from custom ontologies, datasets, and metrics~\cite{buchel2025sok}. Our concern is not with any system's design but with whether the numbers that rank them mean what they appear to.

\subsection{Concurrent Work (2026)}

\vspace{-2mm}

Several systems appearing while this work was in preparation converge on
overlapping goals. GRID~\cite{huang2026grid} trains 4B extractors with task-bank
reinforcement supervision, preserves verbatim evidence anchors, and contributes a
249-article benchmark with an LLM judge calibrated to 86.0\% agreement against
three reviewers. TACTIC-KG~\cite{bouchiha2026tactickg} decomposes construction
into extractor, typer, verifier and curator agents on 3B--8B models.
ANCHOR~\cite{kim2026anchor} enforces SHACL conformance over UCO, STIX~2.1 and
MALOnt on the same 149 reports used here, and TTPrint~\cite{cheng2026ttprint}
anchors each candidate technique to a localized evidence window before verifying
it. Outside security, Loconte et al.~\cite{loconte2026oakmend} formalize
ontology-grounded post-extraction correction as a general recipe.
Related work also considers post-extraction provenance and corroboration
(TRACE-CTI~\cite{valletta2026trace}) and multi-agent CTI graph construction
(KGAgent4CTI~\cite{yang2026kgagent4cti}).

Explicit symbolic validation is therefore no longer distinctive. What remains open, and what this paper measures, is whether the evaluation used to compare any of these systems supports the comparison.

\begin{table}[t]
\centering
\scriptsize
\caption{Matching criteria reported by twelve CTI extraction systems and benchmarks. \checkmark\ marks a rule reimplementable from the publication; $\times$ marks one that is not. Shaded rows are systems also re-scored under eight protocols in Section~\ref{sec:protocol}.}
\label{tab:protocol-survey}
\begin{tabular}{llc}
\toprule
System & Reported matching criterion & Reprod. \\
\midrule
\multicolumn{3}{l}{\textit{Rule specified precisely enough to reimplement}} \\
CTIBench~\cite{alam2024ctibench}      & exact match on named entities & \checkmark \\
\rowcolor{rescored} GRID~\cite{huang2026grid}             & explicit rule set, LLM judge & \checkmark \\
TACTIC-KG~\cite{bouchiha2026tactickg} & embedding similarity, $T{=}0.6$ & \checkmark \\
ANCHOR~\cite{kim2026anchor}           & exact match plus partial credit & \checkmark \\
Fieblinger et al.~\cite{fieblinger2024actionable} & {\sc rouge} $n$-gram overlap & \checkmark \\
\midrule
\multicolumn{3}{l}{\textit{Similarity family named, encoder or threshold unfixed}} \\
\rowcolor{rescored} CTI-Nexus~\cite{cheng2025ctinexus}    & ``semantically match'' (undefined) & $\times$ \\
LLM-TIKG~\cite{hu2024llmtikg}         & cosine similarity, threshold unstated & $\times$ \\
SEvenLLM~\cite{ji2024sevenllm}        & semantic similarity + ROUGE-L & $\times$ \\
\midrule
\multicolumn{3}{l}{\textit{No matching criterion located}} \\
\rowcolor{rescored} CTIKG~\cite{huang2024ctikg}           & not stated & $\times$ \\
\rowcolor{rescored} AttacKG+~\cite{zhang2025attackkg}     & not stated & $\times$ \\
CTIExpert~\cite{zhong2026ctiexpert}   & not stated & $\times$ \\
TTPrint~\cite{cheng2026ttprint}       & not stated & $\times$ \\
\bottomrule
\end{tabular}
\vspace{-2mm}
\end{table}

\vspace{-2mm}
\subsection{Positioning of CTIForge}

\vspace{-2mm}

 Prior work largely optimizes the neural side: CTI-Nexus~\cite{cheng2025ctinexus} improves in-context prompts, CTIExpert~\cite{zhong2026ctiexpert} arbitrates between specialized models, and CTIKG~\cite{huang2024ctikg} orchestrates long-context reasoning across agents. Concurrent 2026 systems add structural control by different means---SHACL conformance in ANCHOR~\cite{kim2026anchor}, a learned verifier in TACTIC-KG~\cite{bouchiha2026tactickg}, span anchoring in TTPrint~\cite{cheng2026ttprint}.

CTIForge occupies a specific point in that space. Extraction is a single standard LLM call, followed by a validation layer that is \emph{deterministic} (the same triple always triggers the same rule), \emph{categorized} (every action maps to one of a fixed taxonomy), and \emph{logged per triple} (the rule that fired is recoverable from the artifact). Among the systems and released artifacts we inspected, none exposes all three properties together. That combination is what makes the error analysis in Section~\ref{sec:error-dist} possible: a learned verifier label alone does not expose which constraint produced it.

\section{Problem Formulation: Scoring as a Free Parameter}
\label{sec:formulation}
\vspace{-2mm}

\vspace{-2mm}

Let $D=\{d_1,\dots,d_n\}$ be CTI documents. A system emits, per document, a set of predicted triples $\hat{Y}_d = \{(e_i, r, e_j)\}$ with $e_i,e_j$ typed entity mentions and $r$ a relation; a benchmark supplies gold triples $Y_d$. Each predicted triple additionally carries an evidence span $s$, a source identifier, and a validation status $\nu \in \{\texttt{raw},\texttt{validated},\texttt{repaired},\texttt{rejected}\}$.

Comparing systems requires deciding when $\hat{y}$ counts as $y$. We make that decision explicit. A \textbf{matching protocol} is a pair $M = (\mu, \sigma)$: a predicate $\mu(\hat{y}, y) \in \{0,1\}$ and a scoping rule $\sigma$ fixing the candidate gold set against which each prediction is tested. Given $M$, true positives follow from a greedy one-to-one assignment in which each gold triple is consumed at most once. Writing the score this way makes explicit that it depends on the protocol as much as on the predictions:

\begin{equation}
\mathrm{F1}(\hat{Y}, Y; M) \;=\; f\bigl(\mu, \sigma, \hat{Y}, Y\bigr).
\end{equation}

Two systems are \textbf{rank-reversed} under $M_1, M_2$ when
$\mathrm{F1}(\hat{Y}^{A};M_1) > \mathrm{F1}(\hat{Y}^{B};M_1)$ and $\mathrm{F1}(\hat{Y}^{A};M_2) < \mathrm{F1}(\hat{Y}^{B};M_2)$.

The predicate $\mu$ varies along three axes used in the literature: \emph{name equality} (exact string, containment, stemmed-token overlap, embedding similarity above a threshold, entailment), \emph{relation equality} (identical after normalization, drawn from a compatibility set, or ignored), and \emph{argument order} (fixed, or swap-tolerant). The scope $\sigma$ is either per-document or pooled over the corpus. Section~\ref{sec:protocol} instantiates eight points in this space and measures the consequences; Appendix~\ref{sec:appendix-protocol} gives each definition in full.

Knowledge-graph quality is a studied problem with an established vocabulary---accuracy, completeness, consistency, timeliness---and a literature on assessing each~\cite{xue2022knowledge}. That vocabulary has not been adopted in CTI extraction, where a single gold-matched F1 stands in for all of it, and the substitution is what we examine. Two properties of a protocol are worth separating. It is \textbf{comparative} only if $\mu$ depends on no participant's ontology---STIX~2.1 vocabulary compliance and evidence presence qualify, a system's own type-pair table does not, since the system that defines the table enforces it at extraction time and cannot score poorly against it. And it is \textbf{calibrated} only to the degree that $\mu$ agrees with a human adjudicating the same pair; Section~\ref{sec:protocol} measures that agreement rather than assuming it.

Our apparatus (Section~\ref{sec:architecture}) targets precision and provenance rather than recall: it rejects or repairs structurally invalid output before graph construction, preserves an evidence span and a validation status on every edge, and merges across documents without opaque coreference. It uses a compact ontology of 14 entity types and 12 relation types; the inventories (Tables~\ref{tab:entity-types}, \ref{tab:relation-types}) and the type-pair constraints (Table~\ref{tab:type-pairs}) are in Appendix~\ref{sec:appendix-ontology}.

\section{Apparatus: The CTIForge Pipeline}
\label{sec:architecture}
\vspace{-2mm}

A design choice cannot be switched off in someone else's released artifact, which is why the audit in Section~\ref{sec:protocol} cannot attribute any component's contribution. CTIForge exists to make that possible. It is a five-stage pipeline (Fig.~\ref{fig:system-overview}) in which each stage is independently toggleable, extraction is separable from everything downstream, and every triple reaching the graph carries a validation status, a source chunk identifier, and an evidence span traceable to the originating sentence.

One property matters for the measurements that follow. Every stage after extraction---validation, saliency filtering, canonicalization, alignment---is deterministic given its input, and link prediction, the only other stage that calls a model, runs \emph{before} the validation toggle. All ablation arms can therefore be derived from one extraction pass, so an arm differs from another by the module under test and by nothing else. Section~\ref{sec:full-run} depends on this.

Extraction is probabilistic; validation is deterministic; canonicalization is conservative and alias-driven rather than embedding-driven; fusion is confidence-aware and provenance-preserving. We describe each stage only to the depth the measurements require; implementation detail is in Appendix~\ref{sec:appendix-validation}.

\subsection{Ingestion and extraction (Modules A--B)}

\vspace{-2mm}
Reports are segmented into paragraph-level chunks with stable identifiers and character offsets back into the source, so every downstream claim retains a pointer to the sentence that produced it. Extraction is a single LLM call per chunk against a schema-constrained prompt, with optional few-shot retrieval and a second call for link prediction only when the document graph is disconnected. Keeping extraction to one call per chunk is what makes the shared-extraction ablation cheap enough to run on seven backbones.

\subsection{Symbolic validation and repair (Module C)}
\label{sec:validation}

\vspace{-2mm}

Module C is the component this paper measures, so we state it precisely. Each predicted triple must pass a deterministic cascade of checks before graph construction: empty-field rejection; indicator-format repair; type repair for \textit{Other}; and type-pair validation against a fixed constraint table, with an auto-swap pass for reversed arguments. Self-loop rejection, evidence alignment against the cited span, placeholder detection for mentions such as ``the attacker'', and ATT\&CK identifier validation complete the sequence. Each check emits one of three actions---\emph{rejected}, \emph{repaired}, or \emph{flagged}---into a per-triple log tagged with one of fourteen error categories. The rule that admitted or altered any edge is therefore recoverable from the artifact rather than from the code.

A separate conservative repair step promotes weak \textit{related\_to} relations to specific types when the evidence supports it, using keyword matching against eleven relation categories. It does not promote \textit{associated\_with} by default and requires both entity mentions to appear in the evidence span before acting.

Two properties of this design are load-bearing later. The checks are hand-written rules with fixed thresholds, developed against the output of a strong extraction model---Section~\ref{sec:full-run} shows what happens when they meet a weaker one. And they are deterministic, which is what allows an ablation arm to be recomputed from stored extraction output rather than re-extracted.

\subsection{Grounding, fusion, and optional stages (Modules D--E)}

\vspace{-2mm}
Module D reduces surface fragmentation in three passes: alias-based canonicalization over a curated table of known actors, malware, and tools; type-gated embedding alignment at a conservative threshold ($\tau=0.85$), so a malware name is never merged with an organization; and ATT\&CK technique grounding, which attaches an identifier as provenance rather than rewriting the surface form. Module E merges per-document graphs while preserving per-triple origin and confidence. Three optional stages---link prediction, embedding entity alignment, and saliency filtering---are retained; three others were implemented and disabled after they over-generated on CTI text (Appendix~\ref{sec:appendix-experiments}).

Because canonicalization rewrites strings that gold annotations record verbatim, it can improve graph usability while costing gold-matched F1. Section~\ref{sec:full-run} measures that trade and finds the metric registers neither gain nor loss.

\section{Experimental Design}
\label{sec:experiments}
\vspace{-2mm}

We first vary the matching protocol to measure evaluation sensitivity, then hold extraction fixed to isolate validation and canonicalization, analyze the resulting error categories, compare systems under uniform criteria, and finally examine backbone and runtime limits.

\subsection{Datasets and Resources}

\vspace{-2mm}
\textbf{CTI-Nexus benchmark.} The primary evaluation corpus consists of 149 gold-annotated CTI reports from the CTI-Nexus dataset~\cite{cheng2025ctinexus} (Table~\ref{tab:data}). Each report contains typed entities and explicit triples in JSON format. An official test split of 11 documents is used for direct comparison with CTI-Nexus reported metrics; the remaining 138 documents serve as the few-shot example pool.

\textbf{CTIKG benchmark.} The secondary benchmark consists of 255 CTI sentences with 693 gold triples from the CTIKG dataset~\cite{huang2024ctikg}. Each sentence is annotated with bracket-format triples and tagged with its ATT\&CK tactic. CTIKG's own predicted triples are included in the dataset, enabling head-to-head comparison under identical evaluation criteria.

\textbf{Train/test separation.} For CTI-Nexus evaluation, few-shot examples are drawn exclusively from the training split. The retriever excludes the current document during example selection to prevent trivial leakage.

\subsection{Research Questions}
\vspace{-2mm}

\begin{enumerate}[leftmargin=*]
\item \textbf{RQ1:} How much of a reported result in this literature is decided by the matching protocol rather than by the system?
\item \textbf{RQ2:} What do symbolic validation (Module C) and canonicalization (Module D) contribute when extraction is held fixed?
\item \textbf{RQ3:} What is the distribution of errors across the symbolic error taxonomy?
\item \textbf{RQ4:} How does CTIForge compare with prior systems under uniform criteria and at matched backbone?
\item \textbf{RQ5:} What limits further improvement, and how much of that limit is the extraction model?
\item \textbf{RQ6:} What does CTIForge cost to run relative to comparable pipelines?
\end{enumerate}

\subsection{Evaluation Metrics}
\label{sec:metrics}

\vspace{-2mm}
Matching is scoped \emph{per document}. Pooling every document's predictions and gold into flat lists before matching permits a prediction from one report to be credited against gold from another; on our data that inflates true positives by 6.4\%. All reported figures use per-document scoping.

We report three families, and keep them separate because they license different claims. \textbf{Gold-matched} metrics (subject--relation--object; relation-free subject--object pair) compare against annotations under a stated matcher. Neither is a strict-equality match: both apply soft name matching, and the triple metric additionally accepts a fixed relation-compatibility set, so both should be read as \emph{soft} protocols throughout. \textbf{Schema-independent} metrics---evidence presence, duplicate rate, and compliance with the external STIX~2.1 relationship vocabulary---depend on no participant's ontology and are therefore valid for cross-system comparison. \textbf{Schema-dependent} metrics scored against our own type-pair table are \emph{definitional}: the validator enforces that table at extraction time, so CTIForge cannot score poorly on them. We report them in Appendix~\ref{sec:appendix-results} as evidence that the layer enforces its contract, never as evidence of superiority over another system.

Matcher definitions, normalization rules, and the type-equivalence groups are given in Appendix~\ref{sec:appendix-protocol}.

\subsection{Experimental Configuration}
\label{sec:experimental-config}

\vspace{-2mm}
All experiments use a precision-first configuration unless otherwise noted. GPT-4o is the default backbone for the full 149-document CTI-Nexus run and the matched-backbone comparison, using OpenRouter~\cite{openrouter} for both systems. The 20-document development slice uses Mistral Small 3.1 24B. The substitution study in Section~\ref{sec:bottleneck} adds MiniMax-M3, Claude Haiku 4.5, and Sonnet 4.5 through hosted APIs, plus Mistral-7B-Instruct, Qwen2.5-7B-Instruct, and Gemma-2-9B offline on one consumer GPU.

All post-extraction stages are identical and deterministic. Hosted extraction uses temperature 0.1; offline extraction is greedy and adds a local JSON-format wrapper to the common prompt. Backbone, serving stack, decoding, and wrapper therefore covary, so this study measures transfer across deployment configurations rather than an isolated serving effect. We excluded two candidate hosted backbones whose temperature settings could not be matched, as discussed in Section~\ref{sec:discussion}.

Validation keeps evidence rejection enabled, canonicalization uses an alignment threshold of $\tau=0.85$, and the retained enhancements are link prediction, embedding-based entity alignment, and saliency filtering. Every configuration is released; Appendix~\ref{sec:appendix-experiments} gives the full hyperparameter table.

\subsection{Ablation Design}
\label{sec:ablation-design}

\vspace{-2mm}
The module-level ablation isolates symbolic validation and canonicalization over the full benchmark. Extraction is run \emph{once} and all arms derive from the same raw triples; every stage downstream is deterministic, so arms differ only by the module under test. Arm names compose cumulatively: B is extraction alone, B+C adds Module C, and B+C+D adds Module D on top of that. Running each arm end-to-end instead would confound the module with a fresh sampling of the extraction model. Five independent extraction passes on MiniMax-M3 at the same temperature put that noise at a standard deviation of 0.0058 in $\Delta$P, against 0.0101 in unvalidated F1: scoring arms from a shared pass removes about half the variance, which is why the design is worth its constraint. The three offline backbones decode greedily and reproduce byte-identically across invocations, so their arms carry no sampling noise at all. The two GPT-4o runs in Tables~\ref{tab:full-results} and~\ref{tab:ablation-full} differ by more, 0.0158 triplet F1 against a validation effect of 0.0169 on that backbone; one was served by the OpenAI API directly and the other routed through OpenRouter, so that figure bounds endpoint and sampling variation together rather than isolating either. We report the retained enhancement configuration only, and do not ablate the optional neuro-symbolic additions around B+C+D; Appendix~\ref{sec:appendix-validation} lists which are enabled.

\section{Results and Analysis}
\label{sec:results}
\vspace{-2mm}

We report the measurement question first. Every subsequent number in this section is produced by a matching protocol, and Section~\ref{sec:protocol} establishes how much that choice decides. The results that follow should be read against that bound, not at face value.

\subsection{RQ1: How much does the matching protocol decide?}
\label{sec:protocol}

\vspace{-2mm}
Every number above is produced by one matching protocol. Two measurements bound how much that choice decides.

\begin{figure}[t]
\centering
\includegraphics[width=0.82\columnwidth]{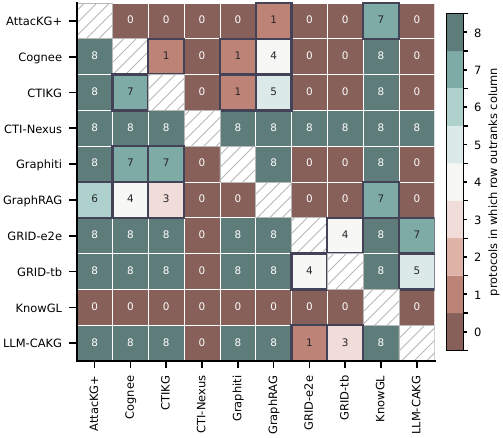}
\caption{Pairwise orderings among ten published systems under eight matching protocols. Each cell counts the protocols in which the row system outranks the column system; shading diverges from the four-four split at which an ordering is unstable. Outlined cells are the eleven pairs whose ordering is not stable across protocols.}
\label{fig:reversal-matrix}
\vspace{-2mm}
\end{figure}

\begin{figure}[t]
\centering
\begin{subfigure}{\columnwidth}
\centering
\includegraphics[width=\columnwidth]{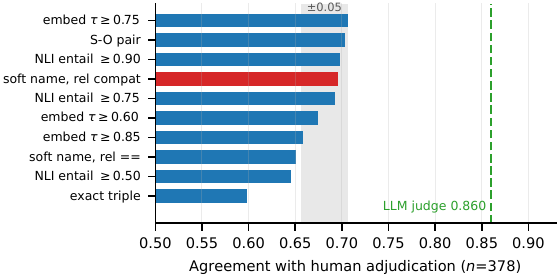}
\caption{Agreement with human adjudication on the 378-item GRID set~\cite{huang2026grid}, ours in red; the band is the $\pm$0.05 resolution at this sample size.}
\label{fig:agreement}
\end{subfigure}

\vspace{0.6em}

\begin{subfigure}{\columnwidth}
\centering
\includegraphics[width=0.80\columnwidth]{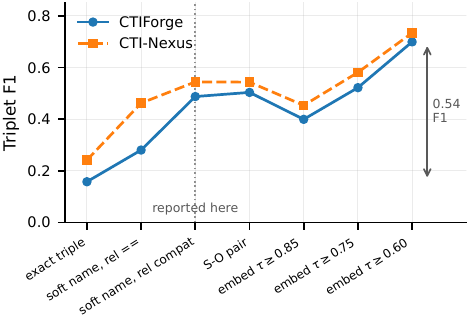}
\caption{Triplet F1 of one fixed prediction set under seven of those protocols, named as in (a). The dotted line marks the protocol we report.}
\label{fig:protocol-sweep}
\end{subfigure}
\caption{The matching protocol as a free parameter. No mechanical protocol reaches the LLM judge, whose figure is the rate at which reviewers agreed with it in~\cite{huang2026grid} rather than an independent re-measurement, and the choice among them moves reported F1 by more than half a point.}
\label{fig:protocol}
\end{figure}

\begin{finding}
\footnotesize \textbf{Finding 1:} Mechanical protocols leave a substantial gap on GRID's human-adjudicated set.
\end{finding}
We score ten protocols---exact match, three soft-lexical variants including ours, three embedding thresholds, and three natural-language-inference thresholds---against the 378 adjudicated items of the GRID judge-calibration set~\cite{huang2026grid}, in which three reviewers ruled on whether an edge matches a reference graph. Agreement spans 0.598 to 0.706 (Fig.~\ref{fig:agreement}), and the dominant failure is \emph{under}-matching: the strict protocol misses 40.2\% of the matches a human credits. Part of that is the matcher being too strict, and part is gold annotation being incomplete in the way Tan et al.~\cite{tan2022revisiting} document for DocRED---a correct prediction with no annotated counterpart is indistinguishable from a wrong one. \emph{No mechanical protocol exceeds 0.71}, including the entailment family, which we added specifically to test whether a semantic matcher escapes the ceiling; it does not. An LLM judge reaches 0.860 on the same items. With 378 items the resolution is about $\pm$0.05, so the mechanical protocols should be read as one group rather than as a ranking; the gap to the judge is several times that width. The instrument the field reports therefore disagrees with human adjudication on roughly three items in ten. We do not read the judge's margin as a recommendation to adopt judges uncritically: they carry documented biases of their own, including position, verbosity, and self-preference effects, and human evaluators are not bias-free either~\cite{chen2024humans}. Nor does a judge's margin transfer across settings. Laskar et al.~\cite{laskar2025improving} benchmark eight LLM judges over three biomedical relation-extraction datasets and find them below 50\% accuracy, largely because extracted relations follow no standard format. That condition holds, notably, for several of the systems in Table~\ref{tab:protocol-survey}. The comparison establishes only that mechanical matchers leave a large gap, not that any particular replacement closes it.

\begin{finding}
\footnotesize \textbf{Finding 2:} A system's rank is substantially a property of the matcher.
\end{finding}
 We score ten published systems---CTI-Nexus, CTIKG, AttacKG+, KnowGL, GraphRAG, Graphiti, Cognee, LLM-CAKG, and two GRID variants---on 50 shared documents under the eight protocols.

\textbf{Provenance of these predictions.} We did not re-run these systems. We use the prediction cache released with GRID~\cite{huang2026grid}. Two criteria motivated that choice. First, \emph{uniform backbone}: every system in the cache was run on the same extraction model, so backbone is held constant and the matcher is the only variable. Second, \emph{shared documents}: all 50 articles are a verified subset of our 149-document benchmark, so the gold annotations are those used throughout this paper. Re-running ten systems ourselves would have reintroduced exactly the confound the audit is designed to remove, since each would have run on whichever backbone its authors assumed.

The inheritance also constrains what the comparison licenses. Because every system ran on a backbone that is not the one its authors selected, these figures are \emph{not} evidence that any pipeline outperforms another, and we do not use them for that; they are evidence only about the sensitivity of orderings to the matcher. Coverage is 50 of our 149 documents, so the reversal count is a lower bound on what a larger shared set might show rather than an estimate of a population value. Systems absent from the cache are absent from this analysis; we did not select them out. Eight protocols over ten systems reverse \textbf{11 of the 45 pairwise orderings} (Fig.~\ref{fig:reversal-matrix}). The swings behind those reversals are not marginal. GraphRAG scores 0.000 under exact match and 0.414 when the relation is ignored. AttacKG+ and KnowGL are likewise zero under one protocol and non-trivial under another, for the same reason: exact string equality is unreachable for a system that emits free-text relations.

\begin{finding}
\footnotesize \textbf{Finding 3:} The same predictions support a fourfold range of reported F1.
\end{finding}
Holding one CTIForge prediction set fixed---the 149-document GPT-4o run of Appendix~\ref{sec:appendix-h2h}---and varying only the matcher across seven protocols moves its triplet F1 (Fig.~\ref{fig:protocol-sweep}) from \textbf{0.1577} under exact match to \textbf{0.6994} under an embedding matcher at $\tau=0.60$, a spread of \textbf{0.5417} on identical predictions (Fig.~\ref{fig:protocol-sweep}). The protocol we report gives 0.4877. CTI-Nexus, scored the same way, spans 0.2416--0.7339. Neither ordering flips in this pair---CTI-Nexus leads under all seven---so here the protocol decides magnitude rather than rank; the ten-system comparison above shows where it decides rank as well. The spread is not caused by any system behaving badly. It follows from three choices that the protocols in Table~\ref{tab:protocol-survey} leave unfixed: whether entity names must be identical or may overlap, whether relation labels must match exactly or merely be compatible, and whether matching is scoped within a document or pooled. Each is a reporting decision rather than an experimental one, and each is invisible in a results table. A reader comparing a published 0.16 against a published 0.70 would infer a different system, not a different set of three decisions.

We place this first because it bounds every comparative claim that follows, including our own. An instrument that disagrees with human adjudication three times in ten on the external calibration set and reorders a quarter of all system pairs in our cache does not support interpreting small differences in gold-matched F1 as evidence about a system. The remaining questions are answered with that limit in view; Section~\ref{sec:discussion} returns to what should be reported instead.

\vspace{-2mm}
\subsection{RQ2: What does the symbolic layer contribute?}
\label{sec:full-run}

\vspace{-2mm}
Table~\ref{tab:full-results} reports the full-pipeline results on all 149 CTI-Nexus-annotated reports (1{,}824 gold triples), using GPT-4o as the extraction backbone and the enhancement flags from Appendix Table~\ref{tab:config}.

\begin{table}[t]
\centering
\scriptsize
\caption{Full-pipeline results on the CTI-Nexus benchmark (149 reports, 1{,}824 gold triples). Triple matching compares subject, relation, and object under the three-pass protocol of Appendix~\ref{sec:appendix-protocol}; entity types are not compared. S-O pair ignores the relation.}
\label{tab:full-results}
\begin{tabular}{lccc}
\toprule
Metric & Precision & Recall & F1 \\
\midrule
Triple (S-R-O)        & 0.5331 & 0.4282 & 0.4749 \\
S-O pair (relation-free) & \textbf{0.5959} & 0.4786 & \textbf{0.5309} \\
Entity type classification & \multicolumn{3}{c}{0.5479 accuracy / 0.5082 micro-F1} \\
\midrule
Predicted triples & \multicolumn{3}{c}{1{,}465} \\
Processing time   & \multicolumn{3}{c}{1{,}586.2\,s total / 10.6\,s per document} \\
\bottomrule
\end{tabular}
\vspace{-2mm}
\end{table}

Two observations structure the rest of the results. First, S-O pair F1 exceeds triple F1 by a wide margin, so the system identifies the participants of most relations and the residual gap is concentrated in relation normalization rather than entity recognition. Second, the pipeline emits fewer candidates than there are gold triples---an intentional consequence of the precision-first configuration. That deficit frames the recall ceiling of Section~\ref{sec:bottleneck}.

\begin{finding}
\footnotesize \textbf{Finding 4:} No tested hosted backbone lost precision under validation; every tested offline backbone did.
\end{finding}
A validation layer is expected to trade recall for precision by discarding bad output. Across the seven tested configurations it does so for all four hosted backbones and for none of the three run offline on our own hardware. Table~\ref{tab:module-ablation} reports three arms per backbone, each derived from one shared extraction pass; the precision column changes sign at that observed boundary, and the groups do not overlap on any column. Because serving mode is not varied independently of backbone, decoding, and backend-specific prompting, the boundary is descriptive rather than causal. It is not ordered by unvalidated F1---the hosted model with the lowest extraction F1 still gains, while an offline model that outscores it still loses---or by parameter count within the offline group, where the largest model falls between the two smaller ones. Seven configurations are an observation, not a law, and two of the hosted changes fall below the 0.0058 pass-to-pass deviation measured in Section~\ref{sec:ablation-design}, assuming comparable variance across hosted backbones.

\begin{figure}[t]
\centering
\includegraphics[width=0.8\columnwidth]{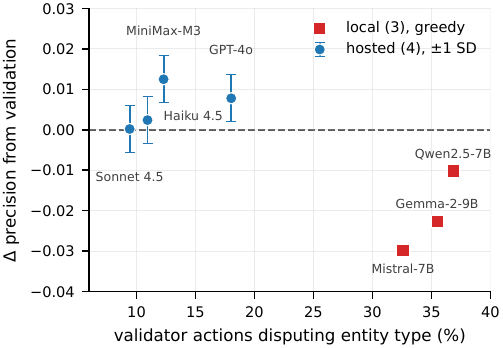}
\caption{Validation-induced precision change against the share of actions disputing entity type, one point per backbone. The dashed line marks no change. Hosted bars are $\pm$1 SD of $\Delta$P over five independent extraction passes on MiniMax-M3; the same interval is drawn on the other three hosted points, for which repeated passes were not run, and the two smallest cross zero. The offline backbones decode greedily and reproduce byte-identically, so they carry no sampling noise. Deployment groups are descriptive because backbone, decoding, and serving stack covary.}
\label{fig:serving-split}
\end{figure}

\begin{finding}
\footnotesize \textbf{Finding 5:} The precision inversion is consistent with a schema-conformance mismatch.
\end{finding}
We classify \emph{type-misassignment} and \emph{impossible-type-pair} actions as those that explicitly dispute entity typing; structural and placeholder actions are kept outside that category. Their share differs sharply between the tested configuration groups (Figure~\ref{fig:serving-split}): roughly one action in eight for hosted backbones and one in three for offline backbones, without overlap. Type-pair constraints are the sharpest instrument in the cascade, since an impossible pair is rejected outright rather than flagged, so a backbone that types its entities loosely can trip the rule on facts that are substantively correct. Our type-pair table, evidence thresholds, and placeholder heuristics were developed against frontier-model output and presume a backbone that respects the ontology it was given. A smaller model can produce triples that are structurally noisier without being less accurate---raw precision on the offline backbones is comparable to the hosted ones, and on two of them higher (Appendix Table~\ref{tab:ablation-full})---and the rules can reject the noise together with the signal. The symbolic half of a neuro-symbolic pipeline is not model-agnostic merely because it is deterministic; it inherits the conventions of whatever produced the text it reads.

\begin{finding}
\footnotesize \textbf{Finding 6:} Where the same weights can be served two ways, serving decides the sign of the validation effect and decoding does not.
\end{finding}
Qwen2.5-7B is the only backbone here available both as local weights and through a hosted endpoint, so it is the only configuration in which the factors of Section~\ref{sec:experimental-config} can be partly separated. Holding weights, prompt, and every downstream stage fixed, we crossed serving with decoding (Appendix~\ref{sec:appendix-2x2}). Both local arms lose precision and both hosted arms gain it: the serving effect is $+0.0214$, about four times the per-pass deviation of Section~\ref{sec:ablation-design}, while decoding and the interaction fall within one. What changes is not how often the validator fires but what it hits---served locally, 55\% of what it removes is correct, against 42\% hosted. One backbone does not generalise the split, but it does show that the deployment stack, not the model, carries it.

\vspace{-2mm}
\begin{finding}
\footnotesize \textbf{Finding 7:} The metric cannot see canonicalization at all.
\end{finding}
B+C+D scores identically to B+C to four decimals on every backbone, yet on MiniMax-M3 the two prediction sets differ on 54 of 1{,}571 triples. Module D rewrites surface forms---alias resolution, casing, ATT\&CK technique names---and the matcher's soft name comparison absorbs every rewrite. The metric is not judging canonicalization to be neutral; it cannot see it.

Section~\ref{sec:protocol} licenses reading these deficits at different weights: under an instrument this protocol-dependent, a third-decimal F1 change is not evidence either way, whereas a sign flip in precision accompanied by a doubling in the share of removals that were correct is.

\begin{table}[t]
\centering
\scriptsize
\setlength{\tabcolsep}{3.5pt}
\caption{Module ablation over seven backbones. Within each backbone, all arms derive from one extraction pass. B is extraction alone; C adds validation and D canonicalization. $\Delta$P is B+C minus B; \emph{collateral} is the share of removals that discard a correct triple, and \emph{type} is the share of actions disputing entity type.}
\label{tab:module-ablation}
\begin{tabular}{llcccrr}
\toprule
Backbone & Serving & B F1 & B+C F1 & $\Delta$P & Collat. & Type \\
\midrule
MiniMax-M3   & hosted & 0.4440 & 0.4371 & $+$0.0125 & 31\% & 12\% \\
GPT-4o       & hosted & 0.5076 & 0.4907 & $+$0.0078 & 46\% & 18\% \\
Haiku 4.5    & hosted & 0.4418 & 0.4255 & $+$0.0024 & 43\% & 11\% \\
Sonnet 4.5   & hosted & 0.4239 & 0.4075 & $+$0.0002 & 47\% & 9\% \\
\midrule
Qwen2.5-7B   & local  & 0.4104 & 0.3706 & $-$0.0102 & 55\% & 37\% \\
Gemma-2-9B   & local  & 0.4305 & 0.3834 & $-$0.0226 & 69\% & 35\% \\
Mistral-7B   & local  & 0.4255 & 0.3663 & $-$0.0298 & 69\% & 33\% \\
\bottomrule
\end{tabular}
\end{table}

\vspace{-2mm}
\subsection{RQ3: How are errors distributed across the taxonomy?}
\label{sec:error-dist}

\vspace{-2mm}
Every validator action is written to a JSONL log tagged with one of the taxonomy categories introduced in Appendix Table~\ref{tab:error-taxonomy}. Table~\ref{tab:error-dist} reports the distribution on both benchmarks, so the profile can be read against the input it was produced from. That profile is configuration-dependent, and the dependence is itself informative rather than a reason to omit it.

\begin{table}[t]
\centering
\scriptsize
\caption{Error taxonomy distribution. Counts are validator actions, not triples; a single triple may generate more than one action. Seven of the fourteen declared categories fire at benchmark scale; the remainder are listed in Appendix Table~\ref{tab:error-taxonomy}.}
\label{tab:error-dist}
\begin{tabular}{lrr}
\toprule
Category & CTI-Nexus (149 docs) & CTIKG (255 sent.) \\
\midrule
Confidence lowered      & 716 (67.0\%) & 173 (31.5\%) \\
Impossible type-pair    & 104 (9.7\%)  & 72 (13.1\%) \\
Repaired identifier     & 104 (9.7\%)  & 23 (4.2\%) \\
\quad\textit{direction swap}       & 58           & 12 \\
\quad\textit{identifier/relation}  & 46           & 11 \\
Type misassignment      & 75 (7.0\%)   & 162 (29.5\%) \\
Generic placeholder     & 38 (3.6\%)   & 96 (17.5\%) \\
Self-loop               & 17 (1.6\%)   & 12 (2.2\%) \\
Hallucinated entity     & 15 (1.4\%)   & 12 (2.2\%) \\
\midrule
Total actions           & 1{,}069 & 550 \\
Sources touched         & 146 / 149 & 193 / 255 \\
\midrule
\multicolumn{3}{l}{\textit{By action}} \\
Flagged                 & 843 (78.9\%) & 443 (80.5\%) \\
Rejected                & 122 (11.4\%) & 84 (15.3\%) \\
Repaired                & 104 (9.7\%)  & 23 (4.2\%) \\
\bottomrule
\end{tabular}
\end{table}

Four findings follow. First, the validator is predominantly a \emph{graded annotator rather than a hard gate}: 78.9\% and 80.5\% of its actions are soft flags, against 11.4\% and 15.3\% outright rejections. Confidence-lowering dominates because the evidence span an LLM cites typically supports a predicted triple partially rather than fully; retaining such triples at reduced confidence preserves graph utility that a binary accept/reject classifier would discard. This is a deliberate design point, and we state it plainly because the term ``validation'' invites the opposite reading.

Second, the profile shifts substantially between benchmarks, and the shift tracks input length. Long CTI-Nexus reports are dominated by confidence-lowering (67.0\%), where partial evidence alignment is the recurring failure. Short CTIKG sentences spread across type misassignment (29.5\%) and generic placeholders (17.5\%), because a single sentence offers little context for typing and frequently refers to entities pronominally. A practitioner porting this validator to a new corpus should expect the dominant category to follow the input granularity rather than remain fixed.

Third, only seven of the fourteen declared categories fire on either benchmark; the remainder guard conditions the parser or alias tables resolve earlier, or are declared but unreached. Appendix Table~\ref{tab:error-taxonomy} gives the full set. We report the observed distribution, not the declared one.

\begin{finding}
\footnotesize \textbf{Finding 8:} The error taxonomy is a census of actions taken, not of errors present.
\end{finding}
An error taxonomy invites reading as a complete accounting of what went wrong. It is not, and the gap runs in both directions. The \textit{repaired identifier} label aggregates three mechanisms, and 58 of its 104 CTI-Nexus actions are subject--object direction swaps rather than identifier normalization (12 of 23 on CTIKG), so we give the split. Type checking is likewise partial by construction: the constraint table rejects pairs that are \emph{impossible}, not merely wrong, flagging 75 misassignments against 600 entities whose type disagrees with gold---roughly one in eight. The layer's guarantee is provenance, not completeness, and a reader taking Table~\ref{tab:error-dist} for a full error census would overestimate both.

\subsection{RQ4: How does CTIForge compare with prior systems?}
\label{sec:ctikg-comparison}

\vspace{-2mm}
Scored under one matcher rather than each system's own, CTIForge and CTIKG sit at opposite points of the same curve: CTIForge is more precise (0.5722 against 0.5122) on roughly 57\% as many triples, CTIKG reaches the higher F1 (0.5900 against 0.5012) on recall. Which point is preferable depends on the downstream task. The instructive number is not either of those: CTIKG's own reported precision, 0.9189, exceeds its uniform-protocol precision by more than 35 points, entirely because its published matcher is more lenient than ours. Appendix~\ref{sec:appendix-ctikg} gives the full comparison.

\textbf{Matched-backbone comparison.} Removing the backbone confound is expected to settle which pipeline is better. It does not, however, and the reason is instructive. Re-running both systems on all 149 documents with the same underlying model leaves CTI-Nexus ahead by 0.057 triplet F1 (0.4877 vs.\ 0.5449), driven entirely by recall---it emits 2{,}092 triples against 1{,}436---while CTIForge holds a 4.4-point precision advantage. Following Section~\ref{sec:protocol} we do not read a gap of that size under a single protocol as a ranking, which is also why the full comparison is reported in Appendix~\ref{sec:appendix-h2h} rather than here. The margins that survive the caveat are the schema-independent ones: under STIX~2.1, an external vocabulary neither system enforces, CTIForge reaches 0.705 relationship compliance against 0.573, and every emitted triple carries a source sentence because the \texttt{Triple} schema requires one, whereas CTI-Nexus emits no evidence field at all. That is a contract difference no choice of matcher affects.

\subsection{RQ5: What limits further improvement?}
\label{sec:bottleneck}

\vspace{-2mm}
Reading the full-pipeline numbers in Table~\ref{tab:full-results} and the same-backbone comparison in Table~\ref{tab:head-to-head} together identifies two concrete bottlenecks separating the current pipeline from higher end-to-end triplet F1, and a third limit set by the extraction backbone itself.

\textbf{Bottleneck 1: relation labeling, not entity identification.} S-O-pair F1 reaches 0.5309 while triplet F1 is 0.4749. The pipeline identifies the participants of a relation more reliably than it labels the relation. Two mechanisms contribute: CTI-Nexus gold uses free-form relation strings that must be normalized into the 12-relation inventory, and a nontrivial fraction collapse into the catch-all types; and semantically adjacent predictions (\textit{uses} vs.\ \textit{exploits}, \textit{delivers} vs.\ \textit{drops}) are not consistently disambiguated. The effect is far larger on CTIKG, whose free-text gold drives strict F1 to 0.0568---a property of the ontology mapping, not of relation quality, and the reason we report the relaxed match as the headline there. The residual gap to CTI-Nexus at matched backbone is likewise recall-dominated, and is the combined effect of saliency filtering, evidence rejection, and cross-chunk deduplication suppressing marginal triples by design.

\textbf{Bottleneck 2: entity typing is the weakest local component.} Accuracy is 0.5479. Errors concentrate on recurring borderline cases---programming languages tagged Tool vs.\ Software, threat-actor names tagged Organization vs.\ ThreatActor---and ontology-guided assignment reports substantially higher figures on this corpus~\cite{kim2026anchor}.

\begin{finding}
\footnotesize \textbf{Finding 9:} Precision is insensitive to the extraction backbone; recall is not.
\end{finding}
A weaker extractor might be expected to degrade the pipeline uniformly, but the two quantities separate across the seven tested backbones (Table~\ref{tab:module-ablation}). Unvalidated precision spans only eight points, with all three offline backbones above the hosted median (Appendix Table~\ref{tab:ablation-full}). Recall separates the groups more sharply: the offline models occupy the bottom and lie within one point of one another.

Figure~\ref{fig:perdoc} puts this difference in proportion. Within any backbone, the interquartile spread of per-document F1 is several times the gap between hosted and offline group means. Recall also varies more widely among hosted models, suggesting that the pipeline's saliency cap, evidence rejection, deduplication, and per-chunk budget become limiting in that regime. The tested free-tier backbone reproduces the paper's qualitative conclusions at no cost. A fully offline configuration remains viable at roughly 0.11 F1 below the strongest tested arm.

\subsection{RQ6: What does the pipeline cost to run?}
\label{sec:efficiency}

\vspace{-2mm}
\begin{finding}
\footnotesize \textbf{Finding 10:} The symbolic layer is not what the pipeline spends its time on.
\end{finding}
On the identical benchmark, backbone, and endpoint, CTIForge completes in roughly two-thirds of CTI-Nexus's wall-clock time. The saving is architectural: CTIForge makes one LLM call per chunk, whereas CTI-Nexus uses a multi-phase prompting pipeline. Validation, canonicalization, and fusion run locally at negligible cost, which also enables the shared-extraction ablation. Appendix~\ref{sec:appendix-cost} gives the timings.

\section{Discussion}
\label{sec:discussion}
\vspace{-2mm}

The strongest supported result is that evaluation protocol changes both score magnitude and, in some comparisons, system ordering; the component findings below should therefore be read as conditional measurements rather than a new universal ranking.

\textbf{What the symbolic layer actually does.} Table~\ref{tab:error-dist} corrects a natural misreading of ``validation'': roughly four in five logged actions are \emph{soft flags}. The layer is therefore predominantly a graded annotator that records why a triple is doubtful, not a gate that discards it. Table~\ref{tab:module-ablation} further shows that its value depends on the output being annotated: the tested hosted configurations trade recall for precision, whereas the tested offline configurations lose precision and mostly remove correct triples.

For analyst-facing graphs, this result favors grading over gating because partially correct triples may retain useful structure. The layer does not guarantee that surviving triples are correct; it guarantees a source sentence, validation status, and rule record for each one. ANCHOR~\cite{kim2026anchor} enforces conformance through SHACL, TACTIC-KG~\cite{bouchiha2026tactickg} through a learned verifier, and CRUcialG~\cite{cheng2025crucialg} through attack-rationality checks that can repair relations. Among the inspected artifacts, we found no equivalent categorized per-triple record; a verifier label alone does not identify the constraint that produced it.

\textbf{Limitations.} At matched backbone, CTIForge trails the recall-oriented CTI-Nexus pipeline by 0.057 triplet F1, with the gap entirely in recall. Raising the saliency cap or relaxing evidence rejection may alter that trade-off, but we do not test those interventions. Mapping free-text gold into 12 relations is also lossy, most visibly on CTIKG (strict F1 0.0568), which is why we report the relaxed match there.

The shared-extraction design isolates downstream modules conditional on one fixed prediction set per backbone; it estimates between-generation variance on one backbone only (Section~\ref{sec:ablation-design}), and the remaining hosted arms assume that estimate transfers. The near-zero Sonnet and Haiku changes are therefore descriptive, not stable effect estimates. Nor do we establish that a soft flag is calibrated: it records which rule fired, not a measured probability of correctness. Finally, ATT\&CK grounding covers only Technique entities---the best open-source LLM reported there reaches 0.22 micro-F1~\cite{ryan2026attackclass}---and cross-document fusion is exercised but not benchmarked. Both require curated evaluation sets that we have not built.

\textbf{Empowering downstream defenses.} Because every edge carries provenance, the graph can be exported to STIX and exchanged without laundering its origin, and validation status supports triage---filter to unrepaired, evidence-aligned edges when confidence matters, admit flagged edges when coverage matters. Evidence that multi-report aggregation improves technique extraction by roughly 26\%~\cite{haque2026beyondsingle} suggests fusion is where the most value remains unrealized.

\textbf{The sampling temperature is becoming unobservable.} Matching decoding settings across hosted endpoints was routine bookkeeping until recently; it no longer is. Several frontier endpoints we screened could not be matched to the 0.1 temperature used by the retained hosted runs: one vendor's current tier rejects the parameter as deprecated, another accepts it but applies it so weakly that repeated calls on an identical prompt reproduced under two-fifths of their own triples, and a third refuses any value but its default. We excluded those hosted backbones rather than report an arm whose sampling condition we could not state. This is the concern of Section~\ref{sec:protocol} one layer lower---there the undeclared free parameter is how predictions are scored, here it is how they were generated---and a configuration file recording a temperature the endpoint did not honor is worse than one recording nothing.

\section{Recommendations}
\label{sec:recommendations}
\vspace{-2mm}

A critique of an instrument is useful only if it says what to do differently. The following recommendations are evidence-based practices suggested by the measured defects, not guarantees that every dataset or system will behave identically. Our results support four changes, none of which requires new annotation.

\subsection{For researchers reporting results}

\vspace{-2mm}
\textbf{R1. Scope matching within a document.} Pooling a corpus into flat prediction and gold lists before matching lets a prediction from one report be credited against a different report's annotations. On our benchmark that inflates true positives by 6.4\%, and nothing in a results table reveals it. Several public harnesses do this. Ours did too, silently, until we audited it.

\textbf{R2. Report a spread, not a point.} A single F1 conceals both that eleven of forty-five system orderings flip under a different protocol and that one fixed prediction set spans 0.54 F1 across protocols (Fig.~\ref{fig:protocol-sweep}). Reporting a protocol spread costs one table; we give ours rather than only prescribing it.

\textbf{R3. Separate metrics by what they license.} Figures scored against a system's own schema are definitional---the system enforces that schema at extraction time and cannot score poorly against it---and cannot rank systems that do not share it. Keeping them apart from schema-independent measures such as STIX~2.1 vocabulary compliance prevents a conformance result from reading as a competitive one. The knowledge-graph community already separates quality dimensions this way~\cite{xue2022knowledge}; the recommendation is to import that discipline, not to invent it.

\textbf{R4. State the matching rule precisely enough to reimplement.} We could reconstruct five of the twelve inspected rules from their publications (Table~\ref{tab:protocol-survey}). ``Semantic matching'' without an encoder and a threshold is not a specification, and a downstream comparison inherits that ambiguity unless it defines a new rule. Releasing an artifact does not substitute for stating the rule: across eleven years of applied security venues, ACSAC reaches 90\% artifact participation in its evaluation committee while only 40\% of those artifacts run~\cite{olszewski2025reproducibility}. Where exact matching is genuinely unsuited to generative output, a multi-dimensional protocol such as GenRES~\cite{jiang2024genres} is preferable to an unspecified similarity, because it can at least be reimplemented.

\subsection{For practitioners selecting tooling}

\vspace{-2mm}
\textbf{R5. Treat a single published F1 as a lower bound on uncertainty, not as a measurement.} The gap between two systems in this literature is routinely smaller than the gap the same system shows across protocols. Where a procurement decision rests on such a comparison, re-score the candidates on your own documents under at least two protocols.

\textbf{R6. Prefer artifacts that carry per-edge provenance.} Whatever the score, an edge that cites the sentence that produced it can be checked by the analyst who must act on it; one that does not, cannot. This property is independent of the matcher and survives any change of protocol.

\section{Conclusion and Future Work}
\vspace{-2mm}

CTI knowledge-graph results are not separable from the protocols that produce them. Re-scoring shows that both reported magnitude and system ordering can change with the matcher, while external human calibration exposes a substantial gap for every mechanical family tested. Matching rules should therefore be explicit, document-scoped, and reported as a spread rather than a single unexplained F1.

The shared-extraction ablation adds a second lesson: deterministic rules did not transfer uniformly across the seven deployment configurations. Their sign split and differing type-action shares are consistent with schema-convention mismatch, but do not isolate serving as its cause. \emph{Deterministic} and \emph{model-agnostic} are different properties.

The external calibration set bounds matcher families rather than measuring CTI triple equivalence in-domain; a multiply annotated CTI set remains necessary. We release the pipeline, configurations, scoring suite, ablation harness, and per-triple records so the reported measurements can be inspected and re-scored.

\section{Ethical Considerations}
\vspace{-2mm}

Our corpora are public vendor reports and their published annotations, and the graphs the pipeline produces contain nothing not already in the source reports. We conducted no experiments involving human subjects: the adjudications used in Section~\ref{sec:protocol} were collected and released by prior work~\cite{huang2026grid} under that work's ethical review, and we reuse them without re-identifying any reviewer. Systems we re-score are evaluated from artifacts their authors released for that purpose.

\bibliographystyle{IEEEtran}
\bibliography{references}

\clearpage
\appendices
\makeatletter
\@addtoreset{table}{section}
\@addtoreset{figure}{section}
\@addtoreset{algorithm}{section}
\makeatother
\renewcommand{\thetable}{\thesection\arabic{table}}
\renewcommand{\thefigure}{\thesection\arabic{figure}}
\renewcommand{\thealgorithm}{\thesection\arabic{algorithm}}

\section{LLM Usage Disclosure}
\label{sec:appendix-llm}
\vspace{-1mm}

We disclose non-minor LLM usage in this work's research pipeline. GPT-4o, accessed via OpenRouter, is the extraction backbone for the main full-benchmark CTIForge run and, in the matched-backbone comparison, for CTI-Nexus as well. The backbone-substitution study evaluates the six additional hosted and offline models listed in Section~\ref{sec:bottleneck}; Mistral Small 3.1 24B is used for the development-slice experiments documented in Appendix~\ref{sec:appendix-experiments}. A sentence-transformer model (all-MiniLM-L6-v2) computes the embeddings used for entity alignment. No LLM was used to generate the claims, analyses, or prose of this paper.

\section{Ontology Details}
\label{sec:appendix-ontology}
\vspace{-1mm}

CTIForge uses a compact ontology of 14 entity types and 12 relation types so that the symbolic schema remains expressive enough for CTI while still being tractable for validation.

\begin{table}[H]
\centering
\scriptsize
\caption{CTIForge entity types.}
\label{tab:entity-types}
\begin{tabular}{p{0.22\linewidth}p{0.68\linewidth}}
\toprule
Entity type & Description \\
\midrule
ThreatActor & Groups, APTs, nation-state actors (e.g., APT29, Lazarus Group) \\
Malware & Malicious software families (e.g., LockBit, Emotet) \\
Tool & Legitimate software misused in attacks (e.g., Cobalt Strike, Mimikatz) \\
Campaign & Named attack campaigns (e.g., Operation Cobalt Kitty) \\
Vulnerability & CVEs and security flaws (e.g., CVE-2023-4966) \\
Technique & ATT\&CK techniques (e.g., T1566 Phishing) \\
Tactic & ATT\&CK tactics (e.g., Initial Access) \\
Organization & Companies, government agencies (e.g., Microsoft, CISA) \\
Location & Countries, regions (e.g., North Korea, Eastern Europe) \\
IOC & Indicators of compromise with subtype: IP, domain, URL, hash, email \\
File & Specific file names, hashes, paths (e.g., payload.dll) \\
Infrastructure & C2 servers, hosting infrastructure \\
Software & Legitimate software products (e.g., Microsoft Exchange) \\
Other & Entities not fitting above categories \\
\bottomrule
\end{tabular}
\end{table}

\begin{table}[H]
\centering
\scriptsize
\caption{CTIForge relation types.}
\label{tab:relation-types}
\begin{tabular}{p{0.26\linewidth}p{0.64\linewidth}}
\toprule
Relation type & Semantic scope \\
\midrule
\textit{uses} & Actor/malware uses tool, technique, or infrastructure \\
\textit{targets} & Actor/campaign targets organization, sector, or location \\
\textit{exploits} & Actor/malware exploits vulnerability or software \\
\textit{delivers} & Campaign/malware delivers payload \\
\textit{communicates\_with} & Malware/tool communicates with IOC or C2 \\
\textit{drops} & Malware drops file or secondary payload \\
\textit{attributed\_to} & Malware/campaign attributed to threat actor \\
\textit{associated\_with} & General association (unconstrained) \\
\textit{variant\_of} & Malware family relationship \\
\textit{located\_in} & Organization/actor located in geography \\
\textit{mitigated\_by} & Vulnerability mitigated by tool, technique, or patch \\
\textit{related\_to} & Catch-all for relations not fitting above (unconstrained) \\
\bottomrule
\end{tabular}
\end{table}

IOC subtypes are intentionally collapsed into one \textit{IOC} type and overlapping relations are collapsed into a smaller inventory so that the validator can constrain outputs without forcing the LLM to discriminate overly fine-grained edge labels.

\section{Validation and Enhancements}
\label{sec:appendix-validation}
\vspace{-1mm}

Algorithm~\ref{alg:validation} gives the full check sequence applied to every raw triple. Order matters: identifier repair and type repair run before type-pair validation, so that a triple is only rejected once its recoverable defects have been corrected. Table~\ref{tab:type-pairs} gives the constraint table the validation step consults.

\begin{algorithm}[t]
\caption{Symbolic Validation Pipeline}
\label{alg:validation}
\begin{algorithmic}[1]
\Require Raw triple $t = (\text{subj}, s_t, \text{rel}, \text{obj}, o_t, \text{ev})$
\Ensure Validated triple $t'$ with status $\nu$
\State reject empty or too-short entities
\State repair identifiers and infer better types for \textit{Other}
\If{$\neg\textsc{ValidTypePair}(s_t, \text{rel}, o_t)$}
  \If{$\textsc{ValidTypePair}(o_t, \text{rel}, s_t)$}
    \State swap arguments and mark as repaired
  \Else
    \State reject
  \EndIf
\EndIf
\State reject forbidden self-loops
\State check evidence alignment and placeholders
\State validate ATT\&CK identifiers
\State assign final status and log taxonomy action
\end{algorithmic}
\end{algorithm}

\begin{table*}[t]
\centering
\scriptsize
\caption{Type-pair constraint table.}
\label{tab:type-pairs}
\begin{tabular}{p{0.14\linewidth}p{0.38\linewidth}p{0.38\linewidth}}
\toprule
Relation & Valid subject types & Valid object types \\
\midrule
\textit{uses} & ThreatActor, Malware, Campaign, Tool & Malware, Tool, Technique, Infrastructure, IOC, Software, File \\
\textit{targets} & ThreatActor, Malware, Campaign, Tool, Vulnerability & Organization, Location, Software, Infrastructure, IOC, File, Vulnerability, Other \\
\textit{exploits} & ThreatActor, Malware, Campaign, Tool, Vulnerability & Vulnerability, Software, Infrastructure \\
\textit{delivers} & ThreatActor, Malware, Campaign & Malware, File, Tool \\
\textit{communicates\_with} & Malware, Tool & IOC, Infrastructure \\
\textit{drops} & Malware, ThreatActor, Campaign & Malware, File, Tool \\
\textit{attributed\_to} & Malware, Campaign, Tool, ThreatActor & ThreatActor, Organization, Location \\
\textit{associated\_with} & \textit{any} & \textit{any} \\
\textit{variant\_of} & Malware & Malware \\
\textit{located\_in} & Organization, ThreatActor, Infrastructure & Location \\
\textit{mitigated\_by} & Vulnerability, Organization, Software & Tool, Technique, Software, Organization \\
\textit{related\_to} & \textit{any} & \textit{any} \\
\bottomrule
\end{tabular}
\end{table*}

The validator also applies regex-based normalization for CVEs, hashes, domains, and ATT\&CK IDs; checks whether evidence text mentions the participating entities; lowers confidence for placeholders such as ``the attacker''; and uses auto-swap repair for reversed but otherwise admissible relations.

\subsection{Optional Neuro-Symbolic Enhancements}
\label{sec:enhancements}

\begin{itemize}[leftmargin=*]
\item \textit{Link prediction}: propose connecting edges between disconnected subgraphs, subject to the same symbolic validator.
\item \textit{Embedding-based entity alignment}: merge same-type mentions with cosine similarity above a conservative threshold.
\item \textit{Saliency filtering}: keep only the highest-scoring triples per document using a symbolic ranker.
\item \textit{Supplement extraction}: a second LLM pass for missed triples; disabled because it over-generated non-gold triples.
\item \textit{Type-constraint-guided re-extraction}: ask about schema-licensed but missing pairs; disabled because valid additions often failed against selective gold annotations.
\item \textit{Multi-signal confidence re-scoring}: recalibrate confidence using symbolic signals; disabled because ranking quality did not align with gold-match status.
\end{itemize}

\section{Extraction Prompt}
\label{sec:appendix-prompt}
\vspace{-1mm}

The extraction prompt is a Jinja2 template with four blocks: an entity-type inventory, a relation-type inventory with disambiguation guidance, a JSON output schema, and fifteen extraction rules. Few-shot examples, when enabled, are appended before the target text. Figure~\ref{lst:prompt} gives its skeleton; the entity and relation inventories are those in Tables~\ref{tab:entity-types} and~\ref{tab:relation-types}.

\begin{figure}[h]
\centering
\begin{promptbox}[Extraction prompt \textnormal{---} \texttt{prompts/extraction.jinja}]
\scriptsize\ttfamily\setlength{\parskip}{2pt}\raggedright
You are a cyber threat intelligence analyst. Extract ALL structured knowledge
triples from the following CTI text.\par
\phead{Entity Types}\par
\pelide{14 types, each with a one-line definition and 3--5 exemplars}\par
\phead{Relation Types}\par
\pelide{12 types; each states directionality and contrasts against the relation
most often confused with it}\par
\phead{Output Format}\par
\{"triples": [\{"subject", "subject\_type", "relation", "object", "object\_type",
"evidence"\}]\}\par
\phead{Rules}\par
\pelide{1. Extract 8--15 triples per chunk, prioritizing core threat relationships.}\par
\pelide{2. Every triple MUST have an evidence span copied from the text.}\par
\pelide{4. NEVER use \texttt{related\_to} if ANY specific relation applies.}\par
\pelide{7. Direction matters: the subject performs the action, the object receives it.}\par
\pelide{15. If no triples can be extracted, return \{"triples": []\}.}\par
\{\% if few\_shot\_examples \%\}\par
\phead{Examples} \pelide{k retrieved exemplars}\par
\{\% endif \%\}\par
\phead{Text to Analyze}\par
\{\{ text \}\}
\end{promptbox}
\caption{Skeleton of the extraction prompt (\texttt{prompts/extraction.jinja}). Grey italics summarize blocks elided for space; rules are numbered as in the source.}
\label{lst:prompt}
\end{figure}

Two design choices in this prompt bear on the results. First, the relation block defines each type \emph{against the type it is most often confused with}, because the dominant strict-match failure is relation-label disagreement rather than entity identification (Section~\ref{sec:bottleneck}). Second, Rule 1 caps output at 8--15 triples per chunk: the extractor is asked to be selective at the point of generation rather than over-generating for a downstream filter to prune, which is consistent with the precision-first operating point and with the ablation result that supplementary extraction passes hurt precision.

\section{Evaluation Protocol}
\label{sec:appendix-protocol}
\vspace{-1mm}

\textbf{Scoping.} All reported metrics match predictions against gold \emph{within a document}. Pooling a corpus into flat prediction and gold lists before matching permits a prediction from one report to be credited against another report's annotations; measured on the 149-document benchmark this inflates true positives by 6.4\% for CTIForge and 5.9\% for CTI-Nexus, corresponding to roughly $+0.03$ F1 for both.

\textbf{Name matching.} Two entity mentions match if, after lowercasing and stripping leading determiners, they are string-equal, one contains the other, or their stemmed token sets overlap by at least 50\% of the shorter set. Agent-noun equivalences (e.g.\ \textit{attacker}\,/\,\textit{attack}) are canonicalized before comparison.

\textbf{Type equivalence.} To avoid penalizing distinctions the gold annotations do not make consistently, three groups are treated as equivalent during type-aware matching: \{Tool, Software, Malware\}, \{Technique, Tactic\}, and \{IOC, File, Infrastructure\}.

\textbf{Relation matching.} The strict matcher requires identical relation labels after normalization into the 12-relation inventory. The relaxed matcher additionally accepts a fixed compatibility set that pairs the catch-all \textit{related\_to} with each specific relation, and admits inverse-perspective pairs such as (\textit{exploits}, \textit{mitigated\_by}) and (\textit{attributed\_to}, \textit{drops}). Matching proceeds in three passes---exact, name-soft with identical relation, then name-soft with compatible relation and a swap-aware check---each pass consuming gold triples greedily so that no gold triple is credited twice.

\textbf{Reproducibility.} Set iteration was replaced by sorted iteration in the subject--object matcher: per-process string-hash randomization otherwise changed greedy match order, producing a $\pm0.0012$ F1 band across runs on identical inputs. All reported figures are deterministic under repeated execution.

\section{Additional Experimental Details}
\vspace{-1mm}

\begin{figure}[t]
\centering
\includegraphics[width=0.8\columnwidth]{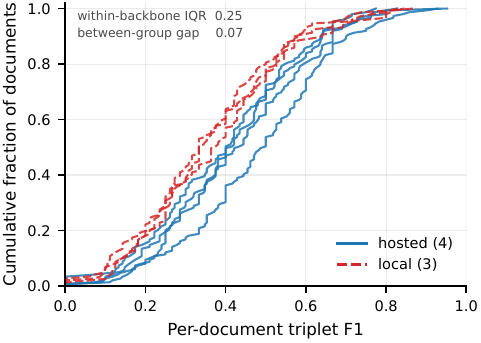}
\caption{Per-document triplet F1 for seven backbones, colored by deployment configuration. Within-backbone spread exceeds the gap between group means, and the groups overlap substantially.}
\label{fig:perdoc}
\end{figure}

This appendix reports the per-backbone results behind Section~\ref{sec:full-run} at full precision. Table~\ref{tab:ablation-full} gives every arm of the module ablation for all seven backbones, so the sign split summarised in the body can be checked column by column; Figure~\ref{fig:perdoc} shows the per-document distribution behind those aggregates.

\begin{table}[t]
\centering
\scriptsize
\caption{Full module ablation across all seven extraction backbones, hosted above the rule and offline below it. Each backbone's arms derive from one shared extraction pass; B is extraction alone, C adds validation and D canonicalization. B+C+D is omitted: it is identical to B+C on every backbone to four decimals, which is the subject of Finding~7. Triples counts predictions emitted. Extraction for this table is a separate pass from Table~\ref{tab:full-results}, so absolute values differ from it by run-to-run sampling; arms within a backbone share one pass and are exactly comparable.}
\label{tab:ablation-full}
\begin{tabular}{llccccr}
\toprule
Backbone & Arm & P & R & Triplet F1 & S-O F1 & Triples \\
\midrule
MiniMax-M3   & B    & 0.4598 & 0.4293 & 0.4440 & 0.4928 & 1{,}703 \\
             & B+C  & 0.4723 & 0.4068 & 0.4371 & 0.4819 & 1{,}571 \\
GPT-4o       & B    & 0.5431 & 0.4764 & 0.5076 & 0.5485 & 1{,}600 \\
             & B+C  & 0.5509 & 0.4424 & 0.4907 & 0.5448 & 1{,}465 \\
Haiku 4.5    & B    & 0.4628 & 0.4227 & 0.4418 & 0.4865 & 1{,}666 \\
             & B+C  & 0.4652 & 0.3920 & 0.4255 & 0.4755 & 1{,}537 \\
Sonnet 4.5   & B    & 0.4777 & 0.3810 & 0.4239 & 0.4654 & 1{,}455 \\
             & B+C  & 0.4779 & 0.3553 & 0.4075 & 0.4604 & 1{,}356 \\
\midrule
Qwen2.5-7B   & B    & 0.4788 & 0.3591 & 0.4104 & 0.4455 & 1{,}368 \\
             & B+C  & 0.4686 & 0.3065 & 0.3706 & 0.4269 & 1{,}193 \\
Gemma-2-9B   & B    & 0.5157 & 0.3695 & 0.4305 & 0.4606 & 1{,}307 \\
             & B+C  & 0.4931 & 0.3136 & 0.3834 & 0.4424 & 1{,}160 \\
Mistral-7B   & B    & 0.5045 & 0.3679 & 0.4255 & 0.4572 & 1{,}330 \\
             & B+C  & 0.4747 & 0.2982 & 0.3663 & 0.4384 & 1{,}146 \\
\bottomrule
\end{tabular}
\end{table}

\section{Data Availability}
\label{sec:appendix-data}
\vspace{-1mm}

We release the code, configurations, matching protocols, and scripts needed to reproduce the reported tables: \url{https://github.com/sbhakim/CTIForge}. The third-party corpora and ten-system prediction cache remain available from their original releases~\cite{huang2026grid}.

\subsection*{Software and hardware environment}

Experiments used Python 3.11. Hosted runs used temperature 0.1 through OpenRouter, while local runs used greedy decoding on an NVIDIA RTX 3090. The released configurations fix software versions and model identifiers: pipeline entity alignment uses \texttt{all-MiniLM-L6-v2}, protocol embeddings use \texttt{all-MiniLM-L12-v2}, and entailment uses \texttt{facebook/bart-large-mnli}. All post-extraction stages are deterministic, including sorted matcher iteration.

\begin{table}[t]
\centering
\scriptsize
\setlength{\tabcolsep}{4pt}
\caption{Datasets and resources.}
\label{tab:data}
\begin{tabular}{p{0.28\linewidth}p{0.10\linewidth}p{0.12\linewidth}p{0.36\linewidth}}
\toprule
Resource & Format & Scale & Role \\
\midrule
CTI-Nexus annotations & JSON & 149 reports & Primary document-level benchmark (1{,}824 gold triples) \\
CTIKG benchmark & CSV & 255 sent. & Secondary sentence-level benchmark (693 gold triples) \\
MITRE ATT\&CK & JSONL & 1{,}076 rec. & Technique grounding and alias lookup \\
CVE database & JSONL & 3{,}011 rec. & Vulnerability identifier validation \\
CWE/CAPEC & JSONL & 1{,}957 rec. & Weakness and attack pattern reference \\
CTI blog corpus & JSONL & 219 articles & Few-shot example retrieval \\
\bottomrule
\end{tabular}
\end{table}

\label{sec:appendix-experiments}

Table~\ref{tab:error-taxonomy} lists the fourteen declared taxonomy categories with their severity and default action; Table~\ref{tab:config} gives the configuration used for every reported run. Both are stated in full so that a run can be reproduced without reading the source.

\begin{table}[t]
\centering
\scriptsize
\setlength{\tabcolsep}{4pt}
\renewcommand{\arraystretch}{1.08}
\caption{Validator taxonomy by category, severity, and default action. Shaded categories are rejected rather than flagged; names match the identifiers in \texttt{error\_taxonomy/error\_log.jsonl}.}
\label{tab:error-taxonomy}
\begin{tabular}{@{}p{0.35\linewidth}@{\hspace{4pt}}l@{\hspace{4pt}}p{0.135\linewidth}@{\hspace{4pt}}p{0.33\linewidth}@{}}
\toprule
Category & Severity & Action & Description \\
\midrule
\rowcolor{sevcrit} \texttt{empty\_field}          & critical & reject        & Empty subject, object, or relation \\
\rowcolor{sevcrit} \texttt{impossible\_type\_pair} & critical & reject        & Violates type-pair constraints \\
\rowcolor{sevcrit} \texttt{self\_loop}            & critical & reject        & Subject equals object \\
\addlinespace[3pt]
\texttt{hallucinated\_entity}  & high   & reject        & Entity absent from evidence \\
\texttt{malformed\_identifier} & high   & repair/reject & Malformed CVE, hash, or ID \\
\texttt{type\_misassignment}   & high   & repair        & Entity type inconsistent with name \\
\texttt{invalid\_relation}     & high   & repair/reject & Relation not in schema \\
\addlinespace[3pt]
\texttt{generic\_placeholder}  & medium & lower conf.   & Generic name (``the attacker'') \\
\texttt{missing\_evidence}     & medium & lower conf.   & Evidence text empty \\
\texttt{duplicate\_triple}     & medium & reject        & Duplicate of accepted triple \\
\addlinespace[3pt]
\texttt{confidence\_lowered}   & low    & lower conf.   & Partial evidence support \\
\texttt{other}                 & low    & log           & Uncategorized issue \\
\addlinespace[3pt]
\texttt{repaired\_identifier}  & info   & repair        & Auto-corrected format \\
\texttt{repaired\_alias}       & info   & repair        & Resolved to canonical name \\
\bottomrule
\end{tabular}
\end{table}

\begin{table}[t]
\centering
\scriptsize
\caption{Impact of configuration on the reported operating point: full hyperparameter set.}
\label{tab:config}
\begin{tabular}{p{0.44\linewidth}p{0.44\linewidth}}
\toprule
Parameter & Value \\
\midrule
Primary LLM (149-doc run) & GPT-4o \\
Same-backbone head-to-head & GPT-4o via OpenRouter \\
Development slice & Mistral Small 3.1 24B via OpenRouter \\
Temperature & 0.1 \\
Max tokens & 4{,}096 \\
Retry attempts & 3 \\
Few-shot examples & 3 \\
Max chunk size & 2{,}000 characters \\
Similarity threshold ($\tau$) & 0.85 \\
Link prediction & enabled \\
Embedding entity alignment & enabled \\
Saliency filter & enabled (max 20 triples/doc) \\
Supplement extraction & disabled \\
Guided re-extraction & disabled \\
Confidence re-scoring & disabled \\
\bottomrule
\end{tabular}
\end{table}

\section{Validator Audit Statistics}
\label{sec:appendix-audit}
\vspace{-1mm}

Table~\ref{tab:audit-stats} reports every validator action logged across both benchmark runs, by category and by the action taken. Counts are actions, not triples: a single triple may trigger more than one rule.

\begin{table}[h]
\centering
\scriptsize
\caption{Impact of each validation rule at benchmark scale: logged actions by category and outcome.}
\label{tab:audit-stats}
\begin{tabular}{lrrr}
\toprule
Category & CTI-Nexus & CTIKG & Default action \\
\midrule
Confidence lowered      & 716 & 173 & flag \\
Impossible type-pair    & 104 & 72  & reject \\
Repaired identifier     & 104 & 23  & repair \\
Type misassignment      & 75  & 162 & flag \\
Generic placeholder     & 38  & 96  & flag \\
Self-loop               & 17  & 12  & reject \\
Hallucinated entity     & 15  & 12  & flag \\
\midrule
\textit{Total actions}  & \textit{1{,}069} & \textit{550} & \\
\textit{Sources touched}& \textit{146 / 149} & \textit{193 / 255} & \\
\midrule
Flagged                 & 843 (78.9\%) & 443 (80.5\%) & \\
Rejected                & 122 (11.4\%) & 84 (15.3\%)  & \\
Repaired                & 104 (9.7\%)  & 23 (4.2\%)   & \\
\bottomrule
\end{tabular}
\end{table}

Seven of the fourteen declared categories fire on either benchmark. Three are implemented, but guard conditions that the output parser or the alias tables resolve before validation is reached: \texttt{empty\_field}, \texttt{malformed\_identifier}, and \texttt{repaired\_alias}. The remaining four are declared but unreached in the current pipeline, and are retained as specification for future work rather than reported as empirical categories: \texttt{invalid\_relation}, \texttt{missing\_evidence}, \texttt{duplicate\_triple}, and \texttt{other}.

\section{Additional Results}
\label{sec:appendix-results}
\vspace{-1mm}

This section collects supplementary results supporting the main claims.

\begin{table}[t]
\centering
\scriptsize
\caption{Head-to-head comparison on the CTIKG benchmark (255 sentences, 693 gold triples) under both a relaxed S-O pair match and a strict triple match. The strict rows show the cost of mapping free-text gold into 12 relations, which is why the relaxed match is the headline here. Numbers in the ``uniform'' rows use the same matcher applied to both systems' outputs; the ``CTIKG (reported)'' row restates the numbers published in the CTIKG paper, which use a different evaluation protocol.}
\label{tab:ctikg}
\begin{tabular}{p{0.32\linewidth}ccccc}
\toprule
System & Precision & Recall & F1 & Predicted \\
\midrule
\multicolumn{5}{l}{\textit{Uniform soft-match evaluation (both systems, same matcher)}} \\
CTIKG               & 0.5122 & \textbf{0.6955} & \textbf{0.5900} & 941 \\
CTIForge (ours) & \textbf{0.5722} & 0.4459 & 0.5012 & 540 \\
\midrule
\multicolumn{5}{l}{\textit{Strict triple match (both systems, same matcher)}} \\
CTIKG               & \textbf{0.3560} & \textbf{0.4834} & \textbf{0.4100} & 941 \\
CTIForge (ours) & 0.0648 & 0.0505 & 0.0568 & 540 \\
\midrule
\multicolumn{5}{l}{\textit{As reported by CTIKG (different protocol, reprinted for reference)}} \\
CTIKG (reported)    & 0.9189 & 0.8939 & $\approx$0.906 & --- \\
\bottomrule
\end{tabular}
\end{table}

\subsection{Schema-Dependent Metrics}

Table~\ref{tab:schema-dependent} reports definitional metrics against CTIForge's own constraint table; they demonstrate contract enforcement, not superiority over a system that did not use that table. The semantic-validity margin is 27.7 points under this internal schema, compared with 13.2 under the external STIX~2.1 vocabulary in Table~\ref{tab:head-to-head}.

\begin{table}[h]
\centering
\scriptsize
\caption{Impact of scoring against the system's own schema (definitional; not a cross-system comparison).}
\label{tab:schema-dependent}
\begin{tabular}{lcc}
\toprule
Metric (scored against CTIForge's constraint table) & CTIForge & CTI-Nexus \\
\midrule
Relation specificity        & 0.809 & 0.532 \\
Entity-type specificity     & 0.980 & 0.795 \\
Semantic validity           & 0.837 & 0.560 \\
Schema constraint compliance& 0.854 & 0.585 \\
Evidence alignment rate     & 0.462 & 0.000 \\
Cross-document entity fusion& 0.107 & 0.085 \\
\bottomrule
\end{tabular}
\end{table}

\section{Matched-Backbone Comparison with CTI-Nexus}
\label{sec:appendix-h2h}
\vspace{-1mm}

Table~\ref{tab:head-to-head} compares both pipelines on all 149 documents using GPT-4o through the same endpoint. Because this comparison uses CTI-Nexus's public matcher, its F1 values are controlled relative-comparison results rather than counterparts to the uniform-matcher values in Table~\ref{tab:full-results}.

\begin{table}[t]
\centering
\scriptsize
\caption{Matched-backbone comparison on 149 documents using GPT-4o via OpenRouter. Gold-matched metrics depend on the stated matcher; schema-independent metrics do not use either system's ontology. Constraint-table metrics appear separately in Appendix~\ref{sec:appendix-results}. Winners are \textbf{bold}.}
\label{tab:head-to-head}
\begin{tabular}{l c c c}
\toprule
Metric & CTIForge & CTI-Nexus & Winner \\
\midrule
\multicolumn{4}{l}{\textit{Gold-matched (per-document scoping)}} \\
Triplet F1                    & 0.4877 & \textbf{0.5449} & CTI-Nexus \\
Triplet precision             & \textbf{0.5536} & 0.5100 & CTIForge \\
Triplet recall                & 0.4359 & \textbf{0.5850} & CTI-Nexus \\
Documents won                 & 57 & \textbf{85} & CTI-Nexus \\
\midrule
\multicolumn{4}{l}{\textit{Schema-independent (comparable across systems)}} \\
STIX 2.1 compliance           & \textbf{0.705} & 0.573 & CTIForge \\
\quad coverage of output      & 0.782 & 0.368 & --- \\
Evidence presence             & \textbf{1.000} & 0.000 & CTIForge \\
Duplicate rate ($\downarrow$) & \textbf{0.022} & 0.055 & CTIForge \\
\midrule
\multicolumn{4}{l}{\textit{Normalization artifacts ($\downarrow$ is cleaner)}} \\
Generic relation rate         & \textbf{0.191} & 0.469 & --- \\
\textit{Other} entity rate    & \textbf{0.021} & 0.205 & --- \\
\midrule
Triples emitted / gold        & 1{,}436 / 1{,}824 & 2{,}092 / 1{,}824 & --- \\
Audit trail                   & 1{,}359 val., 77 rep. & none & CTIForge \\
\bottomrule
\end{tabular}
\end{table}

As discussed in Section~\ref{sec:ctikg-comparison}, the gold-matched rows describe a precision--recall trade-off under one protocol, not a protocol-independent ranking. The STIX~2.1 and evidence-presence rows are schema-independent contract differences.

\section{Runtime Cost}
\label{sec:appendix-cost}
\vspace{-1mm}

On the identical 149-document benchmark, same backbone and same inference endpoint, CTIForge completes in \textbf{26.4\,min} (10.6\,s per document) against CTI-Nexus's 38.7\,min (15.6\,s). Extraction is a single LLM call per chunk, with an optional second call for link prediction only when the document graph is disconnected, whereas CTI-Nexus runs a multi-phase prompting pipeline. The symbolic layer, canonicalization and fusion are deterministic and run locally with no model invocation. Section~\ref{sec:bottleneck} shows the same pipeline runs on a free hosted backbone and offline on one consumer GPU, so cost is a property of the architecture rather than of the model tier.

\section{CTIKG Comparison in Full}
\label{sec:appendix-ctikg}
\vspace{-1mm}

Table~\ref{tab:ctikg} compares CTIForge with CTIKG on the 255-sentence CTIKG benchmark (693 gold triples). The key methodological step is evaluating both systems' outputs with a single matcher (substring containment and token-overlap $\geq 0.5$ after normalization) rather than using each system's own reporting protocol, so that differences in the reported numbers reflect extraction behavior rather than evaluator choice.

The resulting precision--recall trade-off and the gap between uniform and originally reported scores are interpreted in Section~\ref{sec:ctikg-comparison}; the table is retained here to expose the underlying values.

\section{Serving and Decoding Decomposed}
\label{sec:appendix-2x2}
\vspace{-1mm}

Qwen2.5-7B-Instruct is the only backbone in this study available both as local
weights and through a hosted endpoint, so it is the only configuration in which
two of the three covarying factors of Section~\ref{sec:experimental-config} can be
separated. Table~\ref{tab:qwen-2x2} crosses serving with decoding on that
backbone. Weights, prompt content, and every stage after extraction are held
fixed; the local arms differ only in whether generation is greedy, and the hosted
arms only in the requested temperature.

\begin{table}[t]
\centering
\scriptsize
\caption{Qwen2.5-7B under crossed serving and decoding. \emph{Removed} is
B$+$C's reduction in emitted triples; \emph{collateral} is the share of those
removals that discard a correct triple. Both local arms lose precision and both
hosted arms gain it.}
\label{tab:qwen-2x2}
\begin{tabular}{llcccrr}
\toprule
Serving & Decoding & B P & B$+$C P & $\Delta$P & Rem. & Collat. \\
\midrule
local  & greedy  & 0.4788 & 0.4686 & $-$0.0102 & 175 & 54.9\% \\
local  & sampled & 0.4913 & 0.4783 & $-$0.0130 & 163 & 58.9\% \\
\addlinespace[2pt]
hosted & greedy  & 0.4562 & \textbf{0.4703} & $+$0.0141 & 294 & \textbf{42.2\%} \\
hosted & sampled & 0.4496 & \textbf{0.4552} & $+$0.0056 & 289 & \textbf{43.6\%} \\
\bottomrule
\end{tabular}
\end{table}

Averaging over the other factor, the main effect of serving is $+0.0214$ and the
main effect of decoding is $-0.0057$; the interaction is $-0.0057$. Against the
0.0058 per-pass deviation of Section~\ref{sec:ablation-design}, the serving effect
is roughly four deviations and the other two are within one.

Three limits bound this. Each cell is a single pass, so the deviation used to
weigh them is transferred from another backbone rather than measured here.
\emph{Serving} is not remoteness alone: the hosted endpoint may quantise the
weights, applies its own chat template, and enforces JSON through an API
parameter where the local arm uses a prompt wrapper, and it emits roughly a
quarter fewer raw triples. The claim the design supports is that the deployment
stack, taken as a whole, decides the sign---not that any one of its parts does.
And it holds for one backbone; the other six could not be run both ways.

\end{document}